\documentclass[twocolumn,prl,showpacs,preprintnumbers,amsmath,amssymb,superscriptaddress]{revtex4-2}
\usepackage{graphicx}
\usepackage{setspace}

\usepackage{color}
\usepackage[normalem]{ulem}
\usepackage{enumerate}
\usepackage{algorithm}
\usepackage{algpseudocode}
\graphicspath{{./figures/}}

\usepackage{amsmath,amsthm,amssymb,braket}
\usepackage{mathrsfs}

\usepackage[
colorlinks=true, citecolor=blue, urlcolor=blue, linkcolor=blue,
setpagesize=false, bookmarks=false, breaklinks=true
]{hyperref}

\begin{document}
	
	\title{Generalized Keldysh formalism for nonequilibrium correlation functions and \\ its application to fluctuation dynamics}
	\author{Ken Inayoshi}
    \email{kinayoshi@mail.saitama-u.ac.jp}
	\affiliation{Department of Physics, Saitama University, Saitama 338-8570, Japan}
	\author{Hiroshi Shinaoka}
	\affiliation{Department of Physics, Saitama University, Saitama 338-8570, Japan}
	\author{Yuta Murakami}
	\affiliation{Institute for Materials Research, Tohoku University, Sendai 980-8577, Japan}
    \affiliation{RIKEN Center for Emergent Matter Science (CEMS), Wako 351-0198, Japan}

\begin{abstract} 
Recent advances in time-resolved spectroscopies provide increasing access to collective dynamics in correlated quantum materials. 
However, computing the corresponding nonequilibrium two-particle correlation functions remains a major challenge. 
Here, by introducing a contour-dependent virtual probe field within the generalized Keldysh formalism, we propose an approach that computes such correlation functions with the vertex corrections essential for describing collective dynamics.
In particular, we introduce a linear integral equation that computes the correlation functions without explicitly constructing the four-time vertex kernel, and develop its matrix-free Krylov solver based on quantics tensor trains.
Combining our method with nonequilibrium dynamical mean-field theory, we show that the fluctuation dynamics of the order parameter in a nonequilibrium symmetry-broken state depends significantly on whether vertex corrections are included, and that the fluctuation and its decay time grow near the nonthermal critical point.
Our approach thus provides a practical route for evaluating nonequilibrium correlation functions, which are emerging as key observables for characterizing states far from equilibrium.
\end{abstract}

\maketitle

{\it Introduction---}
Driving quantum materials out of equilibrium can generate various transient states and dynamical regimes~\cite{Aoki2014,Giannetti2016,de_la_Torre2021,Murakami2025,Caruso_2026}. 
Their characterization has been advanced by time-resolved spectroscopies. 
In particular, time-resolved Raman scattering~\cite{Kash1985,Yang2017,Versteeg2018,Zhu2019,Han2019,Yang2020,Pellatz2021,Chou2024,Reuveni2026} and resonant inelastic X-ray scattering (RIXS)~\cite{Dean2016,Cao2019,Mitrano2019,Parchenko2020,Mitrano2020,Mazzone2021,Mitrano2024,Merzoni2025,Jost2025,Padma2026} now enable measurements of the real-time evolution of low-energy charge, spin, and phonon excitations.
The observed spectra correspond to nonequilibrium two-particle correlation functions, such as the time-dependent dynamical structure factor.
Furthermore, such correlation functions can provide access to quantum entanglement measures, such as the quantum Fisher information~\cite{Hauke2016,Mathew2020,Scheie2021,Laurell2021,Laurell2025,Scheie2025}, which have recently attracted attention as diagnostics of nonequilibrium states~\cite{Baykusheva2023,Hales2023}.

These experimental developments have stimulated theoretical efforts to compute the corresponding nonequilibrium spectra and correlation functions in correlated electron systems, including time-resolved Raman responses~\cite{Wang2018,Matveev2019,Werner2023,Matveev2026} and RIXS spectra~\cite{Chen2019,Wang2020,Zawadzki2020,Eckstein2021,Werner_2021,Eckstein2026}. 
A direct route is provided by wave-function-based methods, such as exact diagonalization and tensor-network approaches~\cite{Wang2020,Zawadzki2020,Rincon2021}.
However, these approaches are generally limited to small or low-dimensional systems at zero temperature.
As a complementary approach, the nonequilibrium Green's function (NEGF) method~\cite{Kadanoff1962,keldysh1964,Bonitz2016,kamenev2023,stefanucci-Leeuwen2025,Aoki2014} can treat large, high-dimensional systems and naturally incorporate finite-temperature effects.
However, it is based on the one-particle Green's function, which describes one-body excitations as probed by time-resolved angle-resolved photoemission spectroscopy~\cite{Eckstein_arpes2008,Freericks2009,Freericks2021,Sentef2013,Kemper2015,Golez2016,Randi2017,Kemper2017,Kemper2018,Perfetto2020,Eckstein_arpes2021,Schuler_arpes2021,Boschini2024}.
Accurately incorporating many-body correlations to compute nonequilibrium correlation functions requires solving the nonequilibrium Bethe-Salpeter equation (BSE)~\cite{Perfetto2015,stefanucci-Leeuwen2025}, at substantial memory and computational cost~\cite{Aoki2014}.
For this reason, the correlation function is often approximated by the element-wise product of one-particle Green's functions (bubble approximation), neglecting vertex corrections and hence collective effects~\cite{Werner2023}.

\begin{figure*}
    \includegraphics[width=\textwidth]{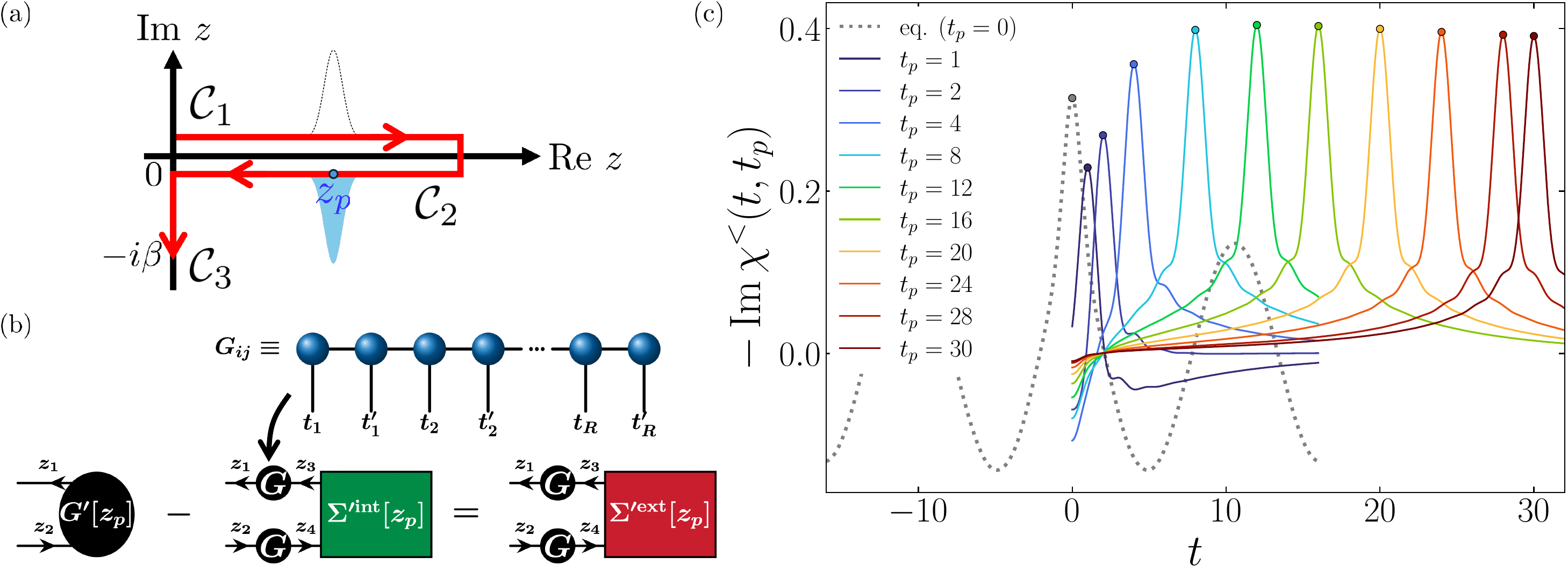}
    \caption{
        (a) Keldysh contour $\mathcal{C}$ and its forward real-time ($\mathcal{C}_1$), backward real-time ($\mathcal{C}_2$), and Matsubara ($\mathcal{C}_3$) branches.
        The blue Gaussian represents the probe field $F(z)$ centered at the probe time $z_p$.
        (b) Diagrammatic expression of the linear integral equation~\eqref{eq:perturbative-GCH}. 
        The tensor network diagram is for $G_{ij}(t,t') \equiv G(t\in \mathcal{C}_i, t'\in \mathcal{C}_j)$, $i,j=1,2,3$.
        (c) Time evolution of the imaginary part of the spin-spin correlation $\chi^<$ (Eq.~\eqref{eq:spin-spin-correlation-function}) for the quench $U=2.0 \to 1.2$.
        The gray dotted line shows the initial equilibrium correlation function, which depends only on $t - t_p$.
        Circles show the fluctuation $\sigma^m(t_p) = -\mathrm{Im}\,\chi^<(t_p, t_p)$ (Eq.~\eqref{eq:order-parameter-fluctuation}).
    }
    \label{fig:keldysh_fluctuation_combined}
\end{figure*}

A promising route beyond this limitation is to introduce a virtual external probe field on the Keldysh contour and compute the correlation function as the functional derivative of a one-particle observable with respect to this field~\cite{Matveev2026}. 
In this work, we develop this idea into a general framework for vertex-corrected nonequilibrium correlation functions using the generalized Keldysh formalism~\cite{Canovi2014} for a generalized contour-dependent Hamiltonian (GCH), whose branch dependence encodes the virtual probe field~\cite{Matveev2026} (Fig.~\ref{fig:keldysh_fluctuation_combined}(a)).
In particular, we introduce a linear integral equation that, given the converged one-particle Green's function, directly computes its variation (Fig.~\ref{fig:keldysh_fluctuation_combined}(b)).
To solve it, we develop a matrix-free Krylov solver that avoids both separate self-consistent simulations for each probe field and the explicit construction of the four-time vertex kernel, representing the two-time objects as quantics tensor trains (QTTs)~\cite{Oseledets2009,Oseledets2010,Khoromskij2011,Shinaoka2023,Murray2024,Sroda2025,Inayoshi2026,Sroda2026,Sroda2026-2}.

As a demonstration, we combine our method with nonequilibrium dynamical mean-field theory (DMFT)~\cite{metzner1989,georges1992,georges1996,schmidt2002,Freericks2006,Eckstein2011,Aoki2014,Murakami2025} and 
study the dynamics of order-parameter fluctuations in a nonequilibrium symmetry-broken state (Fig.~\ref{fig:keldysh_fluctuation_combined}(c)).
We find that the vertex correction drastically alters the fluctuation dynamics, and that the fluctuation is strongly enhanced as the system approaches the nonthermal critical point.

{\it Generalized Keldysh formalism---}
In the conventional Keldysh formalism, the Hamiltonian is identical on the forward and backward branches of the Keldysh contour $\mathcal{C}$, which ensures causality and a simplified implementation~\cite{Schuler2020}. Numerical pump-probe simulations within this framework, however, directly access only the retarded correlation function~\cite{Shao2016,Kumar2019,Kokcu2024,stefanucci-Leeuwen2025}. In equilibrium, this is sufficient because the fluctuation-dissipation theorem allows us to recover the other components. However, this is not the case out of equilibrium. To resolve this,
we add to the Hamiltonian a virtual probe field that depends on $\mathcal{C}$ (Fig.~\ref{fig:keldysh_fluctuation_combined}(a)), and generalize the Kubo formula~\cite{Kubo1957} to access all components of the correlation function (Matsubara, left-mixing, retarded, and lesser).
Namely, we consider a GCH $\hat{H}_F=\hat{H}+F(z)\hat{A}$ and measure an observable $\hat{B}$, where $\hat{H}$ is a time-dependent correlated electron model and $\hat{A}$ is a one-particle operator, such as the spin operator.
$F(z)$ is the virtual field, such as a magnetic field, included on $\mathcal{C}_1$, $\mathcal{C}_2$, or $\mathcal{C}_3$ (see, for example, Fig.~\ref{fig:keldysh_fluctuation_combined}(a)).

Using the generalized Keldysh formalism~\cite{Canovi2014}, the expectation value of $\hat{B}$ for a GCH is defined as
$\langle \hat{B}(z) \rangle= \frac{1}{Z_{\mathcal{C}}}{\rm Tr}\left[T_\mathcal{C} \exp \left\{-i\int_\mathcal{C} d\bar{z} \ \hat{H}_F(\bar{z})\right\}\hat{B}(z)\right]$,
where $z$ is a time variable, $T_\mathcal{C}$ is the time-ordering operator on $\mathcal{C}$, and
$Z_\mathcal{C} \equiv {\rm Tr}\left[T_\mathcal{C} \exp \left\{-i\int_\mathcal{C} d\bar{z} \ \hat{H}_F(\bar{z})\right\}\right]$ is the partition function.
By taking the functional derivative of $\langle\hat{B}\rangle$ with respect to $F$ and taking the limit $F\to 0$,
we can derive the \textit{generalized Kubo formula} for the nonequilibrium correlation function $\chi_{BA}(z,z')\equiv \left. \frac{\delta \langle \hat{B}(z) \rangle}{\delta F(z')}\right|_{F=0}$ as
\begin{align}
    \chi_{BA}(z,z') 
    = i\langle \hat{B}(z) \rangle \langle \hat{A}(z') \rangle -i \langle T_{\mathcal{C}} \hat{B}(z) \hat{A}(z') \rangle. \label{eq:noneq-correlation-function}
\end{align}
Here, we use the functional derivative of the time-evolution operator $\hat{u}_\mathcal{C}(z,z')=T_\mathcal{C} \exp \left\{-i\int^{z}_{z'} d\bar{z} \ \hat{H}_F(\bar{z})\right\}$,
$\frac{\delta \hat{u}_\mathcal{C}(z_1,z_2)}{\delta F(\bar{z})} = -i\hat{u}_\mathcal{C}(z_1,\bar{z})\hat{A}\hat{u}_\mathcal{C}(\bar{z},z_2)$~\cite{Negele1998,Matveev2026}.
The correlation function satisfies
$\delta \langle \hat{B}(z) \rangle = \int_{\mathcal{C}}d\bar{z}\  \chi_{BA}(z,\bar{z})\delta F(\bar{z})$.
Therefore, when we set $\delta F$ as
$\delta F(\bar{z}; z')=\delta h \delta_{\mathcal{C}}(\bar{z},z')$,
where $\delta_{\mathcal{C}}$ is the Dirac delta function on $\mathcal{C}$ and $\delta h$ is an infinitesimal amplitude,
$\delta \langle \hat{B}(z) \rangle$ explicitly depends on $z'$, and
we obtain $\delta \langle \hat{B}(z) \rangle_{z'} = \chi_{BA}(z,z')\delta h$.
Each component of $\chi_{BA}$ can be extracted by choosing the branches of $\mathcal{C}$ on which the probe time $z'$ and the measurement time $z$ lie (see the Supplemental Material~\cite{SM}).

When $\hat{B}$ is a one-particle operator, its response to the probe field can be expressed in terms of the change induced in the one-particle Green's function of the GCH,
$G^F(z,z') = -\frac{i}{Z_{\mathcal{C}}}{\rm Tr}\left[ T_\mathcal{C} \exp \left\{-i\int_\mathcal{C} d\bar{z} \ \hat{H}_F(\bar{z})\right\} \hat{c}(z)\hat{c}^\dagger(z')\right]$,
where $\hat{c}^\dagger$ ($\hat{c}$) is a creation (annihilation) operator.
We denote this change by $\delta G \equiv G^F[\delta h \neq 0] - G^F[\delta h=0]$.
For a GCH, the diagrammatic rules remain unchanged~\cite{Canovi2014}.
Therefore, $G^F$ obeys the nonequilibrium Dyson equation with the self-energy $\Sigma^F(z,z')$~\cite{Canovi2014},
\begin{align}
    G^F(z_1,z_2) = G_0(z_1,z_2) + \left[G_0 * \Sigma^F * G^F\right](z_1,z_2). \label{eq:nonequilibrium-dyson}
\end{align}
Here, $\ast \equiv \int_{\mathcal{C}} d\bar{z}$ is the convolution integral on $\mathcal{C}$.
The self-energy in Eq.~\eqref{eq:nonequilibrium-dyson} consists of two parts, $\Sigma^F = \Sigma^{F,{\rm int}} + \delta \Sigma^{\rm ext}$.
The former collects the interaction diagrams (depending on $F$ through $G^F$), while the latter is the one-particle probe term absorbed as a time-local potential.
$G_0$ is the free Green's function, independent of the probe field.
Because $z$ belongs to one of the three branches, $G^F(z,z')$ has $3 \times 3 = 9$ components.
In the conventional Keldysh formalism, these components are related by causality and reduce to four physical components~\cite{Aoki2014,stefanucci-Leeuwen2025}.
However, in the generalized Keldysh formalism, the probe field breaks causality, and all nine components must be computed~\cite{Canovi2014}.

A straightforward approach to calculating $\delta G$ is to solve Eq.~\eqref{eq:nonequilibrium-dyson} with and without the probe field and take the difference.
A related approach was recently reported for the nonequilibrium nonresonant Raman response of the Falicov-Kimball model within DMFT~\cite{Matveev2026}.
We call this approach the \textit{GCH pump-probe simulation}.
Because the probe field enters the Hamiltonian, $\Sigma^F$ depends on $G^F$, so Eq.~\eqref{eq:nonequilibrium-dyson} must be solved self-consistently for each probe time $z_p$ (see the End Matter for details).
In the linear-response regime, however, $\delta h$ must be small, so $\delta G$ is a tiny difference between nearly identical Green's functions, difficult to extract accurately.
We therefore introduce a linear integral equation for the variation $G^{\prime} \equiv \delta G / \delta h$, which avoids this subtraction.

The perturbed Green's function $G^F$ obeys $G^F = G_0 + \left[G_0 * \left(\Sigma^{F,{\rm int}} + \delta \Sigma^{\rm ext}\right) *G^F\right]$, while the unperturbed one obeys $G=G_0+\left[G_0 * \Sigma^{\rm int} * G\right]$, where $\Sigma^{\rm int}$ is the interaction self-energy without the probe field.
Taking the difference of these two Dyson equations yields the following equation for $G^{\prime} = \delta G/\delta h$ with $\delta G = G^F-G$~\cite{Eckstein2008,stefanucci-Leeuwen2025},
\begin{align}
    &G^{\prime}(z_1,z_2; z_p) - [G * \Sigma^{\prime\rm int}[G, G^{\prime}] *G](z_1,z_2; z_p) \nonumber \\
    &= \mathcal{B}(z_1,z_2; z_p), \label{eq:perturbative-GCH}
\end{align}
where $\Sigma^{\prime\rm int} = \delta \Sigma^{\rm int}/\delta h$ with $\delta \Sigma^{\rm int} = \Sigma^{F,{\rm int}}-\Sigma^{\rm int}$ and $\mathcal{B}(z_1,z_2; z_p) \equiv \left[G* \Sigma^{\prime\rm ext} * G\right](z_1,z_2; z_p)$ with $\Sigma^{\prime\rm ext}=\delta \Sigma^{\rm ext}/\delta h$.
Equation~\eqref{eq:perturbative-GCH} takes the linear form $[\mathcal{A} \star G^{\prime}](z_1,z_2; z_p) = \mathcal{B}(z_1,z_2; z_p)$, where $\star \equiv \int_{\mathcal{C}}dz_5dz_6$ denotes the two-time convolution integral.
The linear superoperator is defined as
$\mathcal{A}(z_1,z_2; z_5, z_6) \equiv \delta_{\mathcal{C}}(z_1, z_5)\delta_{\mathcal{C}}(z_2, z_6)-\int_{\mathcal{C}} dz_3 dz_4 \ G(z_1,z_3) K(z_3,z_4; z_5, z_6) G(z_4,z_2)$.
Here, the kernel $K$ is defined as $K(z_1,z_2; z_3, z_4) \equiv \delta \Sigma^{\rm int}(z_1,z_2)/\delta G(z_3,z_4)$~\cite{stefanucci-Leeuwen2025}
(see Ref.~\cite{SM} for the comparison with the usual BSE).
Both $\mathcal{A}$ and $\mathcal{B}$ depend only on $G$, not on $G^{\prime}$, so Eq.~\eqref{eq:perturbative-GCH} can be solved without self-consistent iteration.
Moreover, unlike in the GCH pump-probe simulation, $G^{\prime}$ is independent of numerical parameters such as the probe pulse width and the field amplitude $\delta h$.

Crucially, Eq.~\eqref{eq:perturbative-GCH} can be solved by a matrix-free Krylov method that requires only the two-time self-energy contractions $G*\Sigma^{\prime\rm int}[G, G^{\prime}]*G$, never the four-time kernel $K$ itself
(see the End Matter for how $\Sigma^{\prime\rm int}$ can be evaluated and for the memory and computational costs).
Often, as in this work, $\Sigma^{\prime\rm int}$ can be supplied as a closed analytic expression.
For generic self-energies evaluated diagrammatically or by quantum Monte Carlo (QMC) simulation~\cite{Rubtsov2005,Werner2006a,Werner2006b,Gull2007,Gull2008,Muhlbacher2008,Werner2009,Schiro2009,Eckstein2009,Werner2010,Gull2011,Eckstein2011},
$\Sigma^{\prime\rm int}$ can be obtained numerically by forward-mode automatic differentiation (the Jacobian-vector product, JVP, method)~\cite{Baydin2018,tenferro-rs}.
We call this approach with Eq.~\eqref{eq:perturbative-GCH} the \textit{perturbative GCH method}.

\textit{Numerical implementation---}
The perturbative GCH method does not depend on how the two-time functions $G$ and $G^{\prime}$ are stored, either as dense matrices on a time grid~\cite{Aoki2014,Schuler2020,NESSi_web,stefanucci-Leeuwen2025} or in compressed form~\cite{Kaye2021,Shinaoka2023,Blommel_phdthesis,Blommel2025,Blommel2026}.
In this work, aiming at future large lattice systems, we adopt the QTT representation~\cite{Oseledets2009,Oseledets2010,Khoromskij2011,Shinaoka2023} (see the tensor network in Fig.~\ref{fig:keldysh_fluctuation_combined}(b)), which gives access to fine time resolution at low memory cost and allows diagrammatic operations such as the element-wise product and the convolution integral to be carried out directly in compressed form~\cite{Shinaoka2023}.
The QTT implementation of Green's functions is explained in Refs.~\cite{Sroda2025,Inayoshi2026,Sroda2026,Sroda2026-2}
(see the End Matter for the bond dimension and truncation cutoff used in the simulation).

We solve Eq.~\eqref{eq:perturbative-GCH} with the generalized minimal residual (GMRES) method~\cite{Saad1986}.
The operator action,
$[\mathcal{A} \star G^{\prime}](z_1,z_2; z_p) = G^{\prime}(z_1,z_2; z_p) - [G * \Sigma^{\prime\rm int}[G, G^{\prime}] *G](z_1,z_2; z_p)$,
requires only sequential contractions of two-time objects.
The GMRES algorithm is given in the End Matter, and
the implementation is available in \texttt{tensor4all-rs}~\cite{tensor4all-rs}, a Rust port of the tensor4all Julia ecosystem~\cite{tensor4all,tensor4all_web}.
We solve Eq.~\eqref{eq:perturbative-GCH} separately for each probe time $z_p$ to obtain $G^{\prime}(z_1, z_2; z_p)$.

\begin{figure*}[t]
    \includegraphics[width=\textwidth]{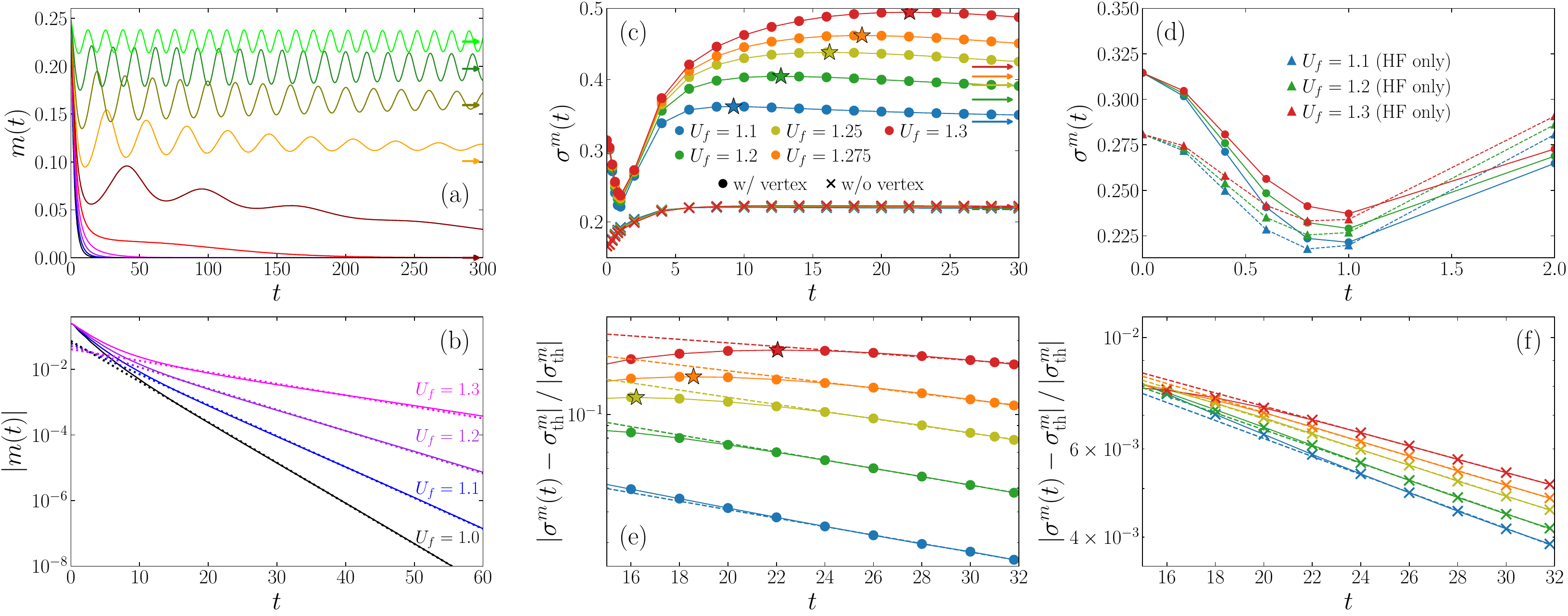}
    \caption{
         Nonequilibrium dynamics of the staggered magnetization $m(t)$ and its fluctuation $\sigma^m(t)$ for quenches $U_i = 2.0 \to U_f$.
         (a) $m(t)$ for $U_f = 1.9, 1.8, \cdots, 1.0$ (from top to bottom).
         The arrows indicate the corresponding thermal values $m_{\rm th}$.
         (b) Logarithmic plot of (a). The dashed lines show fits to exponential relaxations in $t \in [10,50]$.
         (c) $\sigma^m(t)$ with and without the vertex correction.
         The arrows indicate the corresponding thermal values $\sigma^m_{\rm th}$.
         Stars mark the maximum $\sigma^m_{\rm max}$ estimated by parabolic interpolation.
         (d) Comparison of the fluctuation when only the Hartree-Fock term is included in the vertex.
         (e) and (f) Logarithmic plots of $|\sigma^m(t) - \sigma^m_{\rm th}|/|\sigma^m_{\rm th}|$ with and without vertex corrections, respectively.
         The dashed lines are exponential fits in $t \in [24,32)$.
        }
    \label{fig:neq-order-fluctuation}
\end{figure*}

{\it Demonstration---}
Physical quantities encoded in correlation functions, including entanglement measures~\cite{Hauke2016,Laurell2025}, are emerging as promising diagnostics of nonequilibrium states~\cite{Baykusheva2023,Hales2023}. 
We therefore use our approach to study order-parameter fluctuations near a nonequilibrium phase transition, taking as an example an interaction quench from a symmetry-broken state. 
Such quenches exhibit a nonthermal critical point, distinct from the equilibrium critical point, where the transient order-parameter dynamics changes qualitatively~\cite{werner2012,tsuji2013a,tsuji2013b,picano2021,Blommel2025,Blommel_phdthesis}. 
Since previous studies have focused mainly on the order parameter itself, a one-particle observable, it remains unclear whether this nonthermal criticality is also reflected in fluctuations.

We consider a quench of the half-filled Hubbard model on the Bethe lattice~\cite{Eckstein2005,tsuji2013a,tsuji2013b,picano2021},
\begin{align}
    \hat{H}(t) &= -J\sum_{\langle ij \rangle, \sigma}\hat{c}^\dagger_{i\sigma}\hat{c}_{j\sigma} + U(t)\sum_i \left(\hat{n}_{i\uparrow} - \frac{1}{2} \right) \left(\hat{n}_{i\downarrow} - \frac{1}{2} \right), \label{eq:Hubbard-model}
\end{align}
where $\hat{c}^\dagger_{i\sigma}$ creates an electron at site $i$ with spin $\sigma$
and $\hat{n}_{i\sigma} = \hat{c}^\dagger_{i\sigma} \hat{c}_{i\sigma}$.
$-J$ is the nearest-neighbor hopping integral, and we take the infinite-coordination limit with the renormalized hopping $J^*$ fixed.
$U(t)$ is the time-dependent on-site Hubbard interaction.
We set $\hbar =1$ and take $J^*$ and $1/J^*$ as units of energy and time, respectively.

For positive $U$ and sufficiently low temperatures, this system develops antiferromagnetic (AFM) order, whose order parameter is the staggered magnetization $\hat{m} \equiv \hat{S}^z_{\pi}/\sqrt{N}$.
Here, $\hat{S}_i^z = \frac{1}{2}(\hat{n}_{i\uparrow} - \hat{n}_{i\downarrow})$ is the spin operator, 
$N$ is the number of lattice sites, and $\hat{S}^z_{\pi} \equiv \sum_i (-1)^i \hat{S}_i^z/\sqrt{N}$.
The corresponding correlation function is
\begin{align}
    \chi(z, z_p) = -i \left\langle T_{\mathcal{C}} \hat{S}^z_{\pi}(z) \hat{S}^z_{\pi}(z_p)\right\rangle  + i \left\langle \hat{S}^z_{\pi}(z) \right\rangle \left\langle \hat{S}^z_{\pi}(z_p) \right\rangle. \label{eq:spin-spin-correlation-function}
\end{align}
The order-parameter fluctuation is then given by
\begin{align}
    \sigma^m(t) &= -\mathrm{Im}\,\chi^<(t,t) = N\left\{\langle \hat{m}(t) \hat{m}(t) \rangle - \langle \hat{m}(t) \rangle^2\right\}, \label{eq:order-parameter-fluctuation}
\end{align}
which is the main quantity analyzed below.

The time evolution of the model is analyzed within nonequilibrium DMFT~\cite{metzner1989,georges1992,georges1996,schmidt2002,Freericks2006,Eckstein2011,Aoki2014,Murakami2025}.
We approximate the impurity self-energy using second-order perturbation theory, retaining the Hartree-Fock and second-order Born contributions~\cite{tsuji2013b,picano2021}.
To obtain the dynamics of $\sigma^m(t)$ near nonthermal criticality, we solve Eq.~\eqref{eq:perturbative-GCH} for each probe time $t_p$ and compute $\chi^<(t,t_p)$.
The fluctuation at $t_p$ is extracted from the equal-time value, $\sigma^m(t_p)=-\mathrm{Im}\,\chi^<(t_p,t_p)$, as illustrated in Fig.~\ref{fig:keldysh_fluctuation_combined}(c) (see Ref.~\cite{SM} for the implementation).
As a benchmark, we confirm that the results of the perturbative GCH method agree with direct GCH pump-probe simulations at short times and remain stable at longer times (see the End Matter).

We focus on the dynamics after the quench from $U_i = 2.0$ to $U_f$ at $t=0$.
In the initial equilibrium state at inverse temperature $\beta=20$, the system is in the AFM state, characterized by a nonzero staggered magnetization $m=\langle \hat{m} \rangle$~\cite{picano2021}.
After the quench, the closed system is expected to thermalize to an equilibrium state whose effective temperature is fixed by the conserved total energy~\cite{picano2021}.
The thermal values $m_{\rm th}$ and $\sigma^m_{\rm th}$ below are evaluated from equilibrium calculations at $U_f$ and this effective temperature.

\begin{figure}
    \includegraphics[width=0.45\textwidth]{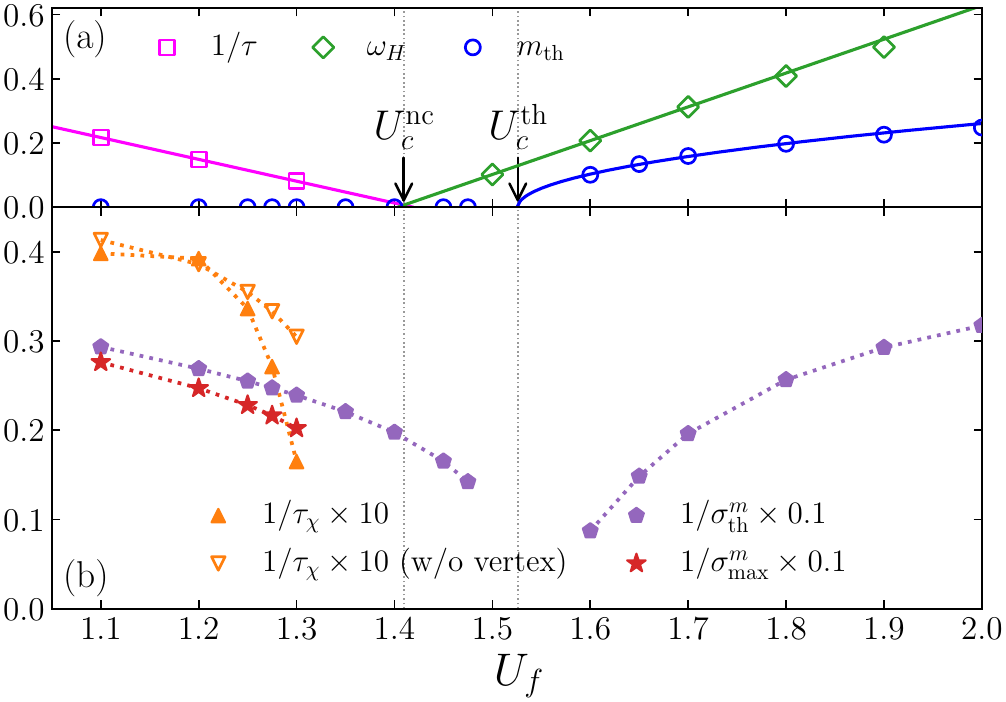}
    \caption{Quantities characterizing the behavior around the critical points $U^{\rm th}_c$ and $U^{\rm nc}_c$.
    (a) The decay time $\tau$ of the order parameter, the Higgs-mode frequency $\omega_H$, and the thermal value $m_{\rm th}$ (see Ref.~\cite{SM} for the estimation of $\omega_H$).
    The thermal critical point $U^{\rm th}_c$ is evaluated from a linear fit to $m_{\rm th}^2$ with the least-squares method.
    $U^{\rm nc}_c$ is estimated from the crossing of the linear fits of $1/\tau$ and $\omega_H$ (solid lines in (a)).
    (b) The decay time $\tau_\chi$ of the fluctuation with and without vertex corrections, and the thermal value $\sigma^m_{\rm th}$ and the maximum $\sigma^m_{\rm max}$ of the vertex-corrected fluctuation (marked in Fig.~\ref{fig:neq-order-fluctuation}(c)).
    The dotted lines in (b) are guides for the eye.
    }
    \label{fig:nonequilibrium-critical}
\end{figure}

We first recall the dynamics of $m(t)$, explored in previous studies~\cite{werner2012,tsuji2013a,tsuji2013b,picano2021}.
For a small quench (e.g., $U_f = 1.9$), $m(t)$ shows coherent oscillations (Higgs mode) around $m_{\rm th}$, whereas for a large quench ($U_f \leq 1.3$) it decays exponentially to $m_{\rm th}$ (Fig.~\ref{fig:neq-order-fluctuation}(a) and~\ref{fig:neq-order-fluctuation}(b)).
As $U_f$ decreases, the center of the oscillation deviates from $m_{\rm th}$, and at $U^{\rm th}_c \approx 1.53$, where $m_{\rm th}$ vanishes as $|U^{\rm th}_c-U_f|^{1/2}$ (blue points in Fig.~\ref{fig:nonequilibrium-critical}(a)), the system reaches the paramagnetic state in the long-time limit but is trapped in a quasi-stable nonthermal fixed point that oscillates around a nonzero value.
As $U_f$ decreases further, this fixed point approaches the nonthermal critical point $U^{\rm nc}_c$, where the Higgs frequency $\omega_H$ vanishes and the relaxation time $\tau$ of $m(t)$ diverges (see Fig.~\ref{fig:nonequilibrium-critical}(a)).

We now turn to the fluctuation dynamics $\sigma^m(t)$, shown in Fig.~\ref{fig:neq-order-fluctuation}(c)--(f).
In the long-time limit, $\sigma^m_{\rm th}$ has a peak at $U^{\rm th}_c$ (see $1/\sigma^m_{\rm th}$, purple points in Fig.~\ref{fig:nonequilibrium-critical}(b)), as in equilibrium (see the End Matter).

In Fig.~\ref{fig:neq-order-fluctuation}(c), we compare the results with and without the vertex correction (see Ref.~\cite{SM} for details).
Immediately after the quench, $\sigma^m$ decreases when the vertex correction is included, but increases when it is ignored.
This first decrease is due to the Hartree-Fock term in the self-energy.
In fact, including only the Hartree-Fock term in the kernel $K$ reproduces this behavior (Fig.~\ref{fig:neq-order-fluctuation}(d), see Ref.~\cite{SM}).
Because the quench decreases $U$, the Stoner enhancement of the susceptibility weakens, and hence $\sigma^m$ decreases
(see the analysis in the End Matter).
This sign reversal implies that vertex corrections are indispensable even at the qualitative level.

After that, $\sigma^m$ increases in both cases, with and without the vertex correction, reaches a peak $\sigma^m_{\rm max}$, and then decreases toward the thermal value $\sigma^m_{\rm th}$ (solid and dashed arrows in Fig.~\ref{fig:neq-order-fluctuation}(c)).
As $U_f$ approaches $U^{\rm nc}_c$, $\sigma^m_{\rm max}$ gradually increases (see the star points in Fig.~\ref{fig:neq-order-fluctuation}(c)).
The decay time of $\sigma^m$ behaves similarly.
In $t \in [24,32)$, exponential fits with and without vertex corrections (Fig.~\ref{fig:neq-order-fluctuation}(e) and (f))
yield the decay time $\tau_\chi$ (orange triangles in Fig.~\ref{fig:nonequilibrium-critical}(b)).
As $U_f$ increases from $1.1$ to $1.3$ toward $U^{\rm nc}_c$, $\tau_\chi$ increases (see $1/\tau_\chi$ in Fig.~\ref{fig:nonequilibrium-critical}(b)).
Interestingly, the difference in $\tau_\chi$ with and without vertex corrections also becomes larger.
Thus, the nonthermal criticality, so far identified only through the order-parameter dynamics, also manifests in the two-particle fluctuation, as an increase in $\sigma^m_{\rm max}$ and $\tau_{\chi}$.

The decay time obtained with the vertex correction appears to approach another critical point different from $U^{\rm nc}_c$.
Clarifying this requires a longer time range than the current limit $t_{\rm max} = 32$, which is set by the slow convergence of GMRES in the long-time region.
A promising route to extend this limit is to suppress the bond dimension of the QTTs, e.g., by the divide-and-conquer method that divides the time domain into segments~\cite{Inayoshi2026,Sroda2026,Grosso2026}, which is left for future work.

\textit{Conclusion---}
We have developed a new approach for computing vertex-corrected nonequilibrium correlation functions using a contour-dependent virtual probe field and a matrix-free QTT solver. 
Applied to nonequilibrium DMFT, the method shows that vertex corrections dramatically reshape antiferromagnetic fluctuation dynamics and that both the fluctuation amplitude and decay time are enhanced near the nonthermal critical point.

The present QTT-based formulation also has several promising extensions. For example, instead of scanning $z_p$, one can treat $G^{\prime}(z_1,z_2;z_p)$ as a single QTT with an additional leg for $z_p$, allowing $G^{\prime}$ to be obtained for many $z_p$ at once. 
The memory efficiency of the QTT representation should be particularly valuable in multi-orbital and large-lattice systems, where the full nine-component storage required by the generalized Keldysh formalism grows rapidly~\cite{Sroda2025,Sroda2026,Sroda2026-2}. 
This makes our approach a promising route for computing nonequilibrium correlation functions in realistic correlated systems.

\textit{Acknowledgments---}
K. I. is grateful to M. Eckstein, P. Werner, M.~{\'S}roda, L. Geng, A. Kauch, A. Ono, N. Tsuji, S. Matsuura, S. Imai, Y. Michishita, and H. Ishida for fruitful discussions.
The DMFT calculation of the order-parameter dynamics in Figs.~\ref{fig:neq-order-fluctuation} and \ref{fig:nonequilibrium-critical} was performed with \texttt{NESSi}~\cite{Schuler2020,Kunzel2026,NESSi_web}.
The one-particle Green's functions were computed with the \texttt{QTT-NEGF} Julia library~\cite{Sroda2025,Inayoshi2026,Sroda2026}, co-developed with M.~{\'S}roda, whom K. I. and H. S. thank for the collaboration.
Our tensor network library \texttt{tensor4all-rs}~\cite{tensor4all-rs,tensor4all,tensor4all_web} used in the calculation of correlation functions is inspired by \texttt{ITensors.jl}~\cite{ITensor}.
Dense tensor operations are executed through its backend
\texttt{tenferro-rs}~\cite{tenferro-rs}, whose strided kernels
(\texttt{strided-rs}~\cite{strided-rs}) are a Rust port of
\texttt{Strided.jl} and \texttt{StridedViews.jl}~\cite{Strided,StridedViews}
employing the {HPTT} tensor-transposition algorithm~\cite{HPTT2017},
with dense linear algebra provided by \texttt{faer-rs}~\cite{faer}.
The quantics time grid on the contour is constructed with
\texttt{quanticsgrids-rs}~\cite{quanticsgrids-rs}, a Rust port of
\texttt{QuanticsGrids.jl}~\cite{tensor4all,tensor4all_web}.
K. I. used the agentic coding tools Claude Code and Cursor for the numerical implementation and the preparation of the manuscript.
All calculations and conclusions were verified by K. I., who takes full responsibility for the content.
K. I. thanks the Supercomputer Center, Institute for Solid State Physics, the University of Tokyo, for the use of the facilities (ISSPkyodo-SC-2026-Ba-0033).
K. I. was supported by JSPS KAKENHI Grant No. 25K17307, Japan.
H. S. was supported by JST FOREST (Grant No. JPMJFR2232) and JSPS KAKENHI Grant No. 23H03817, Japan. 
Y. M. was supported by JSPS KAKENHI Grant Nos. JP24H00191, JP25K07235, JP26K00646, and JP26H01281, Japan.

\bibliographystyle{apsrev4-2}
\bibliography{refs}

\clearpage
\onecolumngrid
\begin{center}
  \textbf{\large End Matter}
\end{center}
\twocolumngrid
\appendix

\textit{Evaluation of $\Sigma^{\prime\rm int}$ and computational cost of the perturbative GCH method---}\label{app:sigma-prime-cost}
As stated in the main text, the perturbative GCH method requires only the action of $K$ on $G^{\prime}$, i.e., the two-time variation $\Sigma^{\prime\rm int}[G, G^{\prime}]$.
In the present demonstration, we use the second-order self-energy, for which $K$ contains two delta functions and could be stored as two-time coefficient functions (see Ref.~\cite{SM}).
The perturbative GCH method becomes more advantageous for self-energies involving internal time integrations, such as resummed diagrammatic approximations and those evaluated numerically, e.g., by QMC~\cite{Rubtsov2005,Werner2006a,Werner2006b,Gull2007,Gull2008,Muhlbacher2008,Werner2009,Schiro2009,Eckstein2009,Werner2010,Gull2011,Eckstein2011}, for which no such sparse structure exists.
The evaluation of such a self-energy is a composition of elementary operations,
$\Sigma^{\rm int}[G] = (f_M \circ \cdots \circ f_1)[G]$,
where each $f_m$ is, e.g., an element-wise product or a convolution.
Writing $X_m = f_m[X_{m-1}]$ with $X_0 = G$, the chain rule gives their variations as $X^{\prime}_m = [\delta f_m/\delta X_{m-1}]\, X^{\prime}_{m-1}$ with $X^{\prime}_0 = G^{\prime}$.
The JVP method~\cite{Baydin2018,tenferro-rs} evaluates this recursion alongside the original computation, applying the differentiation rule registered for each $f_m$ in an automatic differentiation framework, at a memory and computational cost within a constant factor of that of a single evaluation of $\Sigma^{\rm int}$.

For instance, in the $GW$ approximation~\cite{stefanucci-Leeuwen2025}, $K$ becomes a dense four-time function.
Here $\Sigma = iGW$ with the screened interaction $W = v + v * P * W$ ($v$ is the bare interaction) and the polarization $P(z_1,z_2) = -iG(z_1,z_2)G(z_2,z_1)$.
The contribution of $\delta W / \delta G$ renders $K$ dense.
Even in this case, following the above recursion with the intermediates $P$ and $W$,
$W^{\prime}$ and $\Sigma^{\prime}$ can be evaluated with two-time operations only,
$P^{\prime}(z_1,z_2)= -i\left[G^{\prime}(z_1,z_2)G(z_2,z_1) + G(z_1,z_2)G^{\prime}(z_2,z_1)\right]$,
$W^{\prime} = W * P^{\prime} * W$, and
$\Sigma^{\prime} = i\left(G^{\prime}W + GW^{\prime}\right)$.

The computational cost of the perturbative GCH method based on GMRES differs sharply from that of an approach that explicitly constructs the four-time kernel $K$.
As an illustration, suppose that the two-time functions are stored as dense matrices on a time grid with $N_t$ points.
Solving Eq.~\eqref{eq:perturbative-GCH} for a fixed probe time $z_p$ with the kernel formed explicitly would then require $\mathcal{O}(N_t^4)$ memory to store $K$, and each application of $K$ to the two-time function $G^{\prime}$ would cost $\mathcal{O}(N_t^4)$ operations.
In the perturbative GCH method, each GMRES iteration evaluates only the two-time self-energy $\Sigma^{\prime\rm int}$ and the convolutions $G*\Sigma^{\prime\rm int}*G$, which are products of $N_t \times N_t$ matrices and thus require only $\mathcal{O}(N_t^2)$ memory and $\mathcal{O}(N_t^3)$ operations.
In the QTT representation used in this work, the convolution integral is instead carried out as an MPO--MPS contraction, whose computational cost scales as $\mathcal{O}(D_{\rm max}^4)$ and whose runtime memory scales as $\mathcal{O}(D_{\rm max}^3)$ with the maximum bond dimension $D_{\rm max}$~\cite{Sroda2025,Inayoshi2026,Sroda2026,Sroda2026-2}.

\textit{QTT representation and GMRES solver---}\label{app:gmres}
For the QTT calculation, we fix the maximum bond dimension at $D_{\rm max} = 50$--$100$ and truncate each tensor $A$ to $\tilde{A}$ with $\|A-\tilde{A}\|_{\textrm{F}}/\|A\|_{\textrm{F}} \sim 10^{-6}$, where $\|\cdot\|_{\textrm{F}}$ is the Frobenius norm.
This yields a memory footprint far below that of a dense-grid implementation.
Storing the nine components of $G'$ up to $t_{\rm max}=32$ on a dense grid with the real and imaginary time steps $h_t=h_\tau=0.02$ would require $0.29$~GB, whereas our QTT calculation uses only $\mathcal{O}(10^{-3})$~GB while resolving much finer time structure ($h_t=2.8\times10^{-14}$, $h_\tau = 5.7\times10^{-13}$)~\cite{Inayoshi2026}.

The overall procedure to solve Eq.~\eqref{eq:perturbative-GCH} is summarized in Algorithm~\ref{alg:linear-response}.
Given the converged Green's function $G$ and the perturbation $\Sigma^{\prime\rm ext}$, we compute the right-hand side $\mathcal{B}$ and solve the linear equation $\mathcal{A} \star G^{\prime} = \mathcal{B}$ by GMRES.
Each application of the operator $\mathcal{A} \star G^{\prime}$ (the function \texttt{ApplyOperator}) evaluates the linearized interaction self-energy $s = \Sigma^{\prime\rm int}[G, G^{\prime}]$, obtained analytically in this work (see Ref.~\cite{SM}) or via the JVP, and the sandwich convolution $G * s * G$.
We adopt the convergence criterion $\|\mathcal{B} - \mathcal{A}\star G^{\prime}\|_{\textrm{F}}/\|\mathcal{B}\|_{\textrm{F}} < \epsilon_{\rm tol}$, with $\epsilon_{\rm tol} \sim \mathcal{O}(10^{-5})$--$\mathcal{O}(10^{-4})$ in this work.
The Frobenius norms here include a sum over all $3\times3=9$ contour components.
Since Eq.~\eqref{eq:perturbative-GCH} is a linear equation with fixed $\mathcal{B}$, this residual-based criterion directly reflects the accuracy of the obtained solution.

\begin{algorithm}[H]
\caption{Solving Eq.~\eqref{eq:perturbative-GCH} via GMRES with QTT}
\begin{algorithmic}[1]
\State \textbf{Input:} Converged Green's function $G$, perturbation $\Sigma^{\prime\rm ext}$.
\State \textbf{Function} $\Sigma^{\rm int}[G]$: Computes the interaction self-energy.

\State \textbf{Step 1: Compute $\mathcal{B}$}
\State $\mathcal{B} \leftarrow G * \Sigma^{\prime\rm ext} * G$ \Comment{MPO contraction with truncation}
\State \textbf{Step 2: Define Linear Operator $\mathcal{A} \star G^{\prime}$}
\Function{ApplyOperator}{$G^{\prime}$}
    \State // Compute interaction part
    \State $s \leftarrow \Sigma^{\prime\rm int}[G, G^{\prime}]$ \Comment{Evaluated analytically (this work) or via JVP}
    \State $y \leftarrow G * s * G$ \Comment{Sandwich convolution}
    \State $y \leftarrow \text{Truncate}(y, D_{\rm max})$ \Comment{Compress MPO bond dimension}
    \State \Return $G^{\prime} - y$
\EndFunction

\State \textbf{Step 3: Solve Linear Equation}
\State $G^{\prime} \leftarrow \text{GMRES}(\text{ApplyOperator}, \mathcal{B})$

\State \textbf{Output:} $G^{\prime}$
\end{algorithmic}
\label{alg:linear-response}
\end{algorithm}

In this work, we use the restarted GMRES, GMRES($m$)~\cite{Saad1986}, which bounds the Krylov subspace dimension to $m$ and keeps the memory and orthogonalization costs of the QTT basis vectors low.
In each outer restart cycle $k$, we compute the true residual $r^{(k)} = \mathcal{B} - \mathcal{A}\star G^{\prime\,(k)}$ and run an inner cycle of at most $m$ steps that approximately solves the correction equation $\mathcal{A}\star \eta = r^{(k)}$ from a zero initial guess, updating the solution as $G^{\prime\,(k+1)} \leftarrow G^{\prime\,(k)} + \eta$.
A complication specific to the QTT representation is that the bond-dimension truncation applied after each Arnoldi step, i.e., each extension of the Krylov basis by a new vector, degrades the orthonormality of the basis.
To compensate, we reorthogonalize each new basis vector against the existing ones within the inner cycle~\cite{tensor4all-rs}.
We set $m \le 10$ and perform several tens of iterations to obtain the converged solution.

\textit{Details of the GCH pump-probe simulation and benchmark---}\label{app:benchmark}
Because the probe field breaks causality, the standard causal time-stepping algorithm for the Kadanoff-Baym equations~\cite{Aoki2014,stefanucci-Leeuwen2025} is not applicable to the GCH pump-probe simulation.
Instead, the self-consistent solution of Eq.~\eqref{eq:nonequilibrium-dyson} must be obtained by a global iteration on the entire two-time domain for all nine contour components~\cite{Canovi2014}.
Moreover, since the correlation function is obtained as $\chi = \delta G / \delta h$ with $\delta G = G^F - G$, $G^F$ and $G$ must be converged to within the target error of $\chi$ times the small amplitude $\delta h$.

\begin{figure}
    \includegraphics[width=0.45\textwidth]{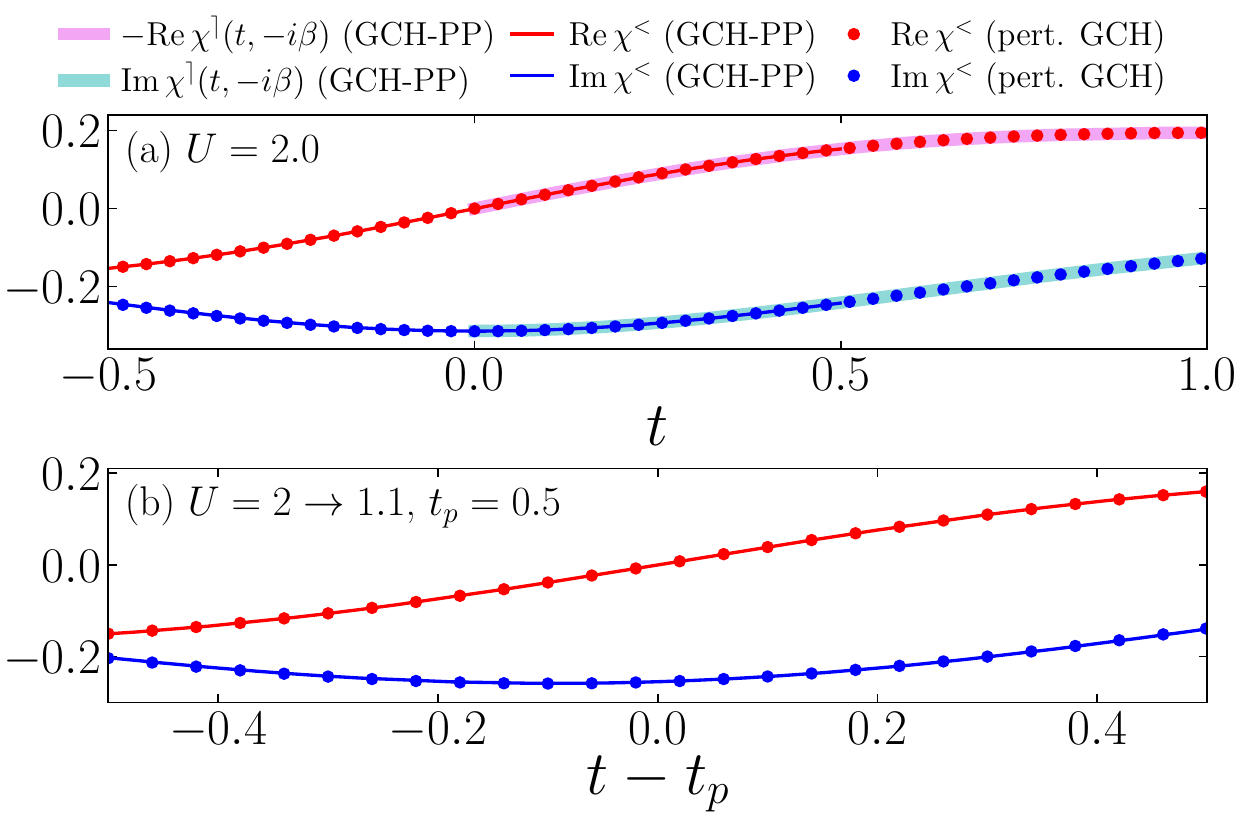}
    \caption{
        Comparison of the spin-spin correlation function $\chi(z, z_p)$ calculated by the GCH pump-probe simulation (GCH-PP, solid lines) and the perturbative GCH method (pert. GCH, points).
        (a) Equilibrium case for $U=2$ and $\beta = 20$, where we plot $\chi^<(t)$ since $\chi^<$ depends only on the relative time $t$.
        The thick magenta and cyan lines show the left-mixing component $\chi^\rceil(t, -i\beta)$.
        (b) Nonequilibrium case for $U_i = 2.0 \to U_f = 1.1$ and $\beta = 20$, where we plot $\chi^<(t, t_p)$ at the fixed probe time $t_p = 0.5$ as a function of $t-t_p$.
        }
    \label{fig:compare-dyson-bse}
\end{figure}

We benchmark the perturbative GCH method against the GCH pump-probe simulation by comparing the spin-spin correlation function $\chi(z, z_p)$.
In the GCH pump-probe simulation, we approximate the delta-function probe pulse by a narrow Gaussian $(\delta h/(\sqrt{2\pi}w))\exp[-(t-t_p)^2/{2w^2}]$ with width $w = 10^{-10}$ and amplitude $\delta h = 10^{-3}$, constructed in the QTT form by the tensor cross interpolation (TCI)~\cite{Oseledets201070,Nunez2022,Ritter2024,tensor4all,tensor4all_web}.
In the perturbative GCH method, neither $w$ nor $\delta h$ is needed, since the right-hand side of Eq.~\eqref{eq:perturbative-GCH} is evaluated directly with the delta-function probe pulse.

Figure~\ref{fig:compare-dyson-bse}(a) compares the lesser spin-spin correlation function $\chi^<(t)$ obtained from the GCH pump-probe simulation (solid lines) and from the perturbative GCH method (points) in the equilibrium case ($U=2$, $\beta = 20$) over the short-time range up to $t_{\rm max} = 1.0$, and the two results agree well.
We also confirm the Kubo-Martin-Schwinger (KMS) relation $-(\chi^<(t))^* = \chi^\rceil(t, -i\beta)$, equivalent to the fluctuation-dissipation theorem~\cite{stefanucci-Leeuwen2025}.
The left-mixing component $\chi^\rceil(t, -i\beta)$ (thick magenta and cyan lines in Fig.~\ref{fig:compare-dyson-bse}) is obtained in the GCH pump-probe simulation by injecting the virtual probe pulse at $z_p = -i\beta \in \mathcal{C}_3$ (see Ref.~\cite{SM}).
However, for longer times $t_{\rm max} \gtrsim 1$, the global iteration becomes harder to converge to the accuracy required for extracting the small difference $\delta G$, whereas the perturbative GCH method remains stable.
The same comparison in the nonequilibrium case ($U_i = 2.0 \to U_f = 1.1$, $\beta = 20$) also shows good agreement (Fig.~\ref{fig:compare-dyson-bse}(b)).

\begin{figure}[htbp]
    \includegraphics[width=0.45\textwidth]{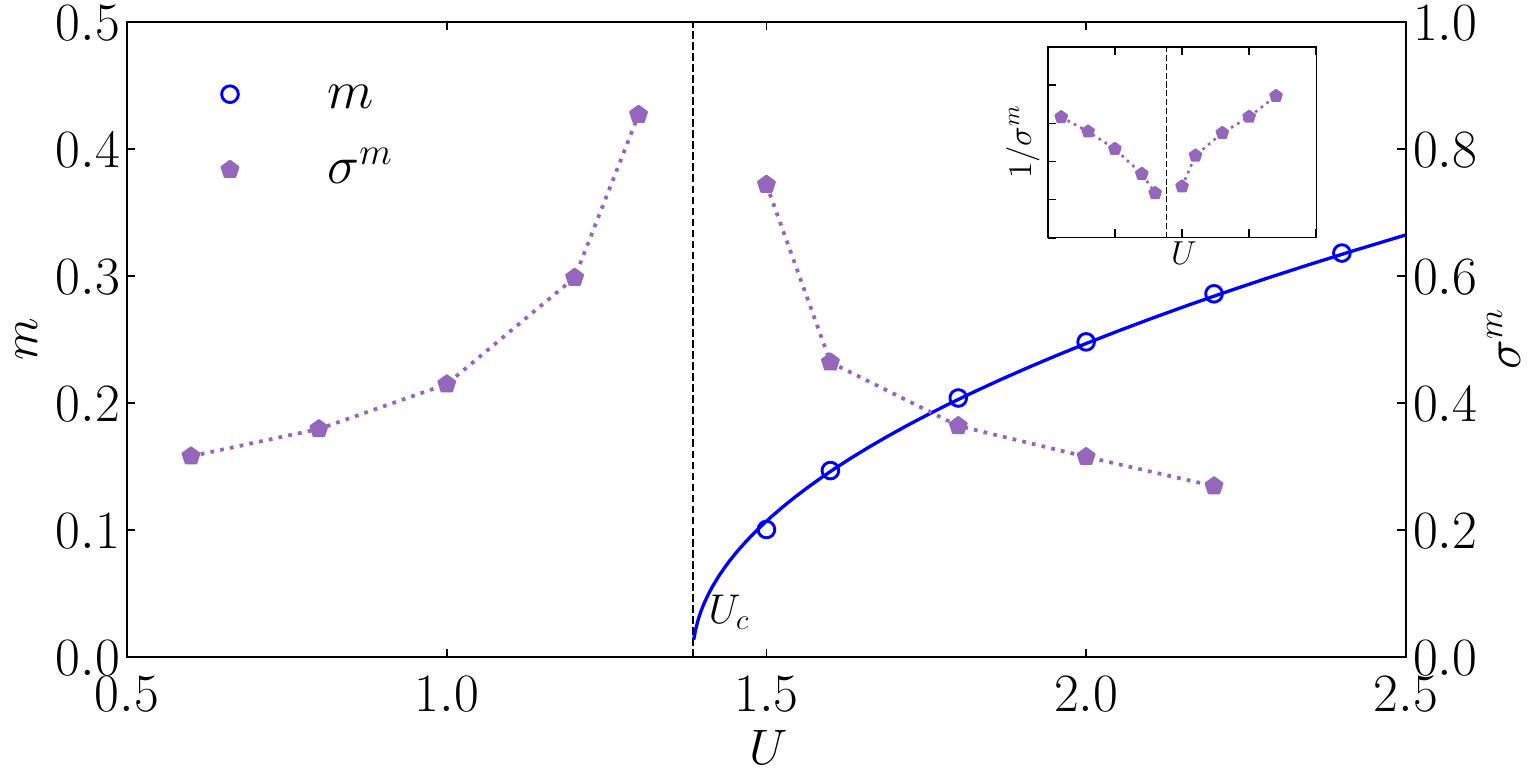}
    \caption{
        Staggered magnetization $m$ and its fluctuation $\sigma^m$ at $\beta = 20.0$. The inset shows the inverse of $\sigma^m$.
        The vertical black dashed line shows the critical point $U_{\rm c}\approx 1.38$.
        }
    \label{fig:equilibrium-results}
\end{figure}

\textit{Equilibrium order parameter and its fluctuation---}
Figure~\ref{fig:equilibrium-results} shows the equilibrium staggered magnetization $m$ and its fluctuation $\sigma^m$ at $\beta = 20.0$.
As the Hubbard interaction $U$ is decreased,
the staggered magnetization $m$ decreases as $|U_{\rm c}-U|^{1/2}$ near the critical point $U_{\rm c}\approx 1.38$.
As expected, the fluctuation $\sigma^m$ has a peak at $U_{\rm c}$.

\textit{Analysis using the random phase approximation---}\label{app:rpa}
We interpret the initial decrease of $\sigma^m$ after the quench using the equilibrium random phase approximation (RPA).
In equilibrium, using the fluctuation-dissipation theorem, we can express $\sigma^m=-\mathrm{Im}\,\chi^<(0)$ as
\begin{align}
    \sigma^m
    &\propto -\int_0^{\infty} \coth{\left(\frac{\beta\omega}{2}\right)} \mathrm{Im}\,\chi^R(\omega) d\omega, \nonumber \\
    &= \int_0^{\infty} \coth{\left(\frac{\beta\omega}{2}\right)}
    \frac{b(\omega)}{(1-2Ua(\omega))^2 + (2Ub(\omega))^2} d\omega.
\end{align}
Here, we use $\chi^R(\omega)=\chi_0^R(\omega)/(1+2U\chi_0^R(\omega))$ and express the bare susceptibility as $\chi_0^R(\omega) = - a(\omega) -i b(\omega)$ with the real functions $a$ and $b$.
Because the hyperbolic cotangent $\coth{\left(\frac{\beta\omega}{2}\right)}$ diverges as $\omega\to 0$, the integral is dominated by the region around $\omega=0$.
In this region, the Lehmann representation guarantees $a(\omega)>0$ and $b(\omega)\geq 0$
($a(0)>0$ and $b(0)=0$).
Thus, the factor $(1-2Ua(\omega))$ in the denominator plays an important role.
Since the Green's function does not immediately deviate from its equilibrium form after the quench, the early-time effect of the quench is a reduction of $U$ in the denominator, which enlarges $(1-2Ua(\omega))^2$ and hence reduces $\sigma^m$, explaining the numerically observed initial decrease.

%
%

\clearpage

\onecolumngrid
\raggedbottom
\begin{center}
  \textbf{\large Supplemental Material for}\\[2pt]
  \textbf{\large ``Generalized Keldysh formalism for nonequilibrium correlation functions and its application to fluctuation dynamics''}\\[6pt]
  {\normalsize by Ken Inayoshi, Hiroshi Shinaoka, and Yuta Murakami}
\end{center}

\setcounter{page}{1}
\renewcommand{\thepage}{SM-\arabic{page}}
\setcounter{section}{0}
\setcounter{equation}{0}
\setcounter{figure}{0}
\setcounter{table}{0}
\renewcommand{\thesection}{S\arabic{section}}
\renewcommand{\theequation}{S\arabic{equation}}
\renewcommand{\thefigure}{S\arabic{figure}}
\renewcommand{\thetable}{S\arabic{table}}

\section{Components of the nonequilibrium correlation function}\label{sm:noneq-correlation-function}
\begin{figure*}[hbt]
  \includegraphics[width=\textwidth]{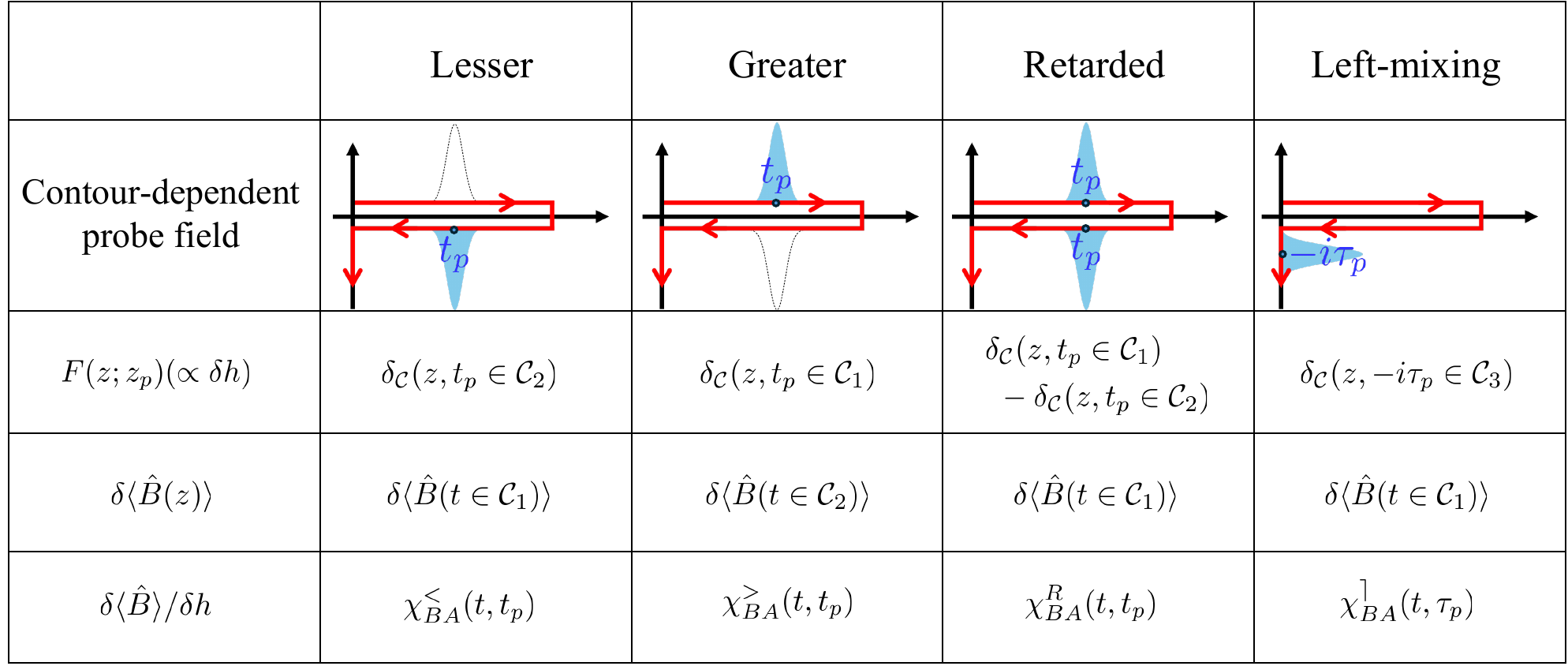}
  \caption{
    Relation between the contour-dependent probe field $F(z; z_p)$, the measured response $\delta \langle \hat{B} \rangle$, and the corresponding component of the nonequilibrium correlation function $\chi_{BA}$ for each position of the probe time $z_p$ on the contour.
  }
  \label{fig:PP-table}
\end{figure*}
Figure~\ref{fig:PP-table} summarizes the relation between the contour-dependent probe field~\cite{Matveev2026} and the components of the nonequilibrium correlation function for the probe time $z_p$.
When the virtual probe field is injected at $z_p = t_p \in \mathcal{C}_2$ ($\mathcal{C}_1$) and the change in the observable $\hat{B}$ is evaluated at $t \in \mathcal{C}_1$ ($\mathcal{C}_2$), i.e., $\delta \langle \hat{B}(t \in \mathcal{C}_1) \rangle$ ($\delta \langle \hat{B}(t \in \mathcal{C}_2) \rangle$),
the lesser (greater) component of the correlation function $\chi_{BA}^<(t, t_p)$ ($\chi_{BA}^>(t, t_p)$) is obtained.
When we consider the usual physical probe field, which takes the same value on the two real-time branches
(i.e., the physical pulse $\delta(t, t_p)$ is written as $\delta_{\mathcal{C}}(z, t_p \in \mathcal{C}_1) - \delta_{\mathcal{C}}(z, t_p \in \mathcal{C}_2)$),
the retarded component $\chi_{BA}^R(t, t_p)$ is obtained from $\delta \langle \hat{B}(t \in \mathcal{C}_1) \rangle$.
This is consistent with the usual Kubo formula~\cite{Kubo1957}.
Finally, when the virtual probe field is injected at $z_p = -i\tau_p \in \mathcal{C}_3$ and $\delta \langle \hat{B}(t \in \mathcal{C}_1) \rangle$ is evaluated,
the left-mixing component $\chi_{BA}^\rceil(t, \tau_p)$ is obtained.

\section{Comparison of Eq.~\eqref{eq:perturbative-GCH} with the usual BSE}\label{sm:comparison-with-usual-bse}
\begin{figure*}
    \includegraphics[width=\textwidth]{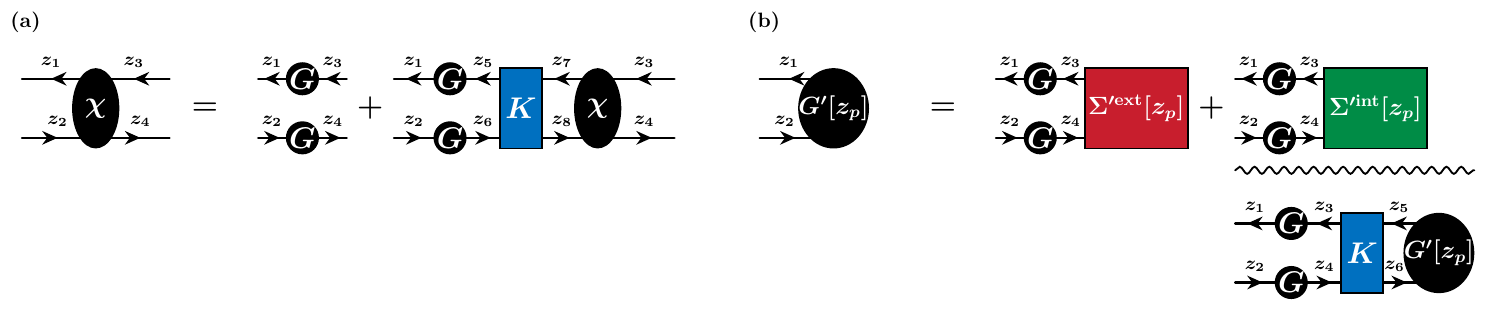}
    \caption{
        Comparison of the diagrammatic representation of 
        (a) the usual BSE and
        (b) Eq.~\eqref{eq:perturbative-GCH}.
        }
    \label{fig:compare-bse-and-perturbative-gch}
\end{figure*}
Figure~\ref{fig:compare-bse-and-perturbative-gch} compares the diagrammatic representations of the usual BSE and Eq.~\eqref{eq:perturbative-GCH}.
Written as a linear equation using the vertex kernel $K$, Eq.~\eqref{eq:perturbative-GCH} has the same algebraic structure as the usual BSE for the four-time two-particle Green's function, with the two external time arguments contracted with the localized probe vertex at $z_p$.
In fact, an equation similar to Eq.~\eqref{eq:perturbative-GCH} appears in the derivation of the usual BSE~\cite{stefanucci-Leeuwen2025}.
As mentioned in the main text, in this work we solve Eq.~\eqref{eq:perturbative-GCH} in terms of two-time functions only, without storing the kernel $K$ as a four-time object.

\section{Nonequilibrium DMFT setup}\label{sm:noneq-dmft}
In this work, we add an external staggered magnetic field as a probe to the Hamiltonian studied in the main text (Eq.~\eqref{eq:Hubbard-model}) and consider the following Hamiltonian,
\begin{align}
    \hat{H}_F(z; z_p) &= -J\sum_{\langle ij \rangle, \sigma}\hat{c}^\dagger_{i\sigma}\hat{c}_{j\sigma} + U(t)\sum_i \left(\hat{n}_{i\uparrow} - \frac{1}{2} \right) \left(\hat{n}_{i\downarrow} - \frac{1}{2} \right) -\sum_i h_i(z; z_p)\hat{S}_i^z.
\end{align}
$h_i(z; z_p)$ is the external magnetic field at site $i$, and we consider a delta-function magnetic field at time $z_p$, i.e., $h_i(z; z_p) = \delta h \delta_{\mathcal{C}}(z,z_p)(-1)^{i}$, where $\delta h$ is the infinitesimal amplitude and $(-1)^i$ is $1$ ($-1$) on the A (B) sublattice~\cite{Eckstein2005}.

In DMFT~\cite{metzner1989,georges1992,georges1996,schmidt2002,Freericks2006,Eckstein2011,Aoki2014,Murakami2025}, 
the lattice self-energy is assumed to be local, and the local self-energy and the local lattice Green's function are identified with those of the effective impurity model, i.e.,
$\Sigma_{ij}^{\mathrm{latt}} = \delta_{ij}\Sigma^{\mathrm{imp}}$ and $G_{ii}^{\mathrm{latt}} = G^{\mathrm{imp}}$, which becomes exact in the infinite-dimensional limit. 
We define the local Green's function in DMFT as
\begin{align}
  G^F_{\sigma}(z,z'; z_p) = -i\langle T_{\mathcal{C}} \hat{c}_{\sigma}(z) \hat{c}_{\sigma}^{\dagger}(z')\rangle_{z_p},
\end{align}
where $\langle \cdots \rangle_{z_p}$ denotes the expectation value with respect to the Hamiltonian that depends on the probe time $z_p$.
This Green's function satisfies the nonequilibrium Dyson equation for the effective impurity model,
\begin{align}
  G^F_{\sigma}(z,z'; z_p) = g_0(z,z') +\left[g_0 * \left(\Delta^F_\sigma + \Sigma^{F,{\rm int}}_{\sigma} + \delta\Sigma_{\sigma}^{\rm ext}\right)* G^F_{\sigma} \right] (z,z'; z_p),
\end{align}
where $g^{-1}_0(z,z') = i\partial_z \delta_{\mathcal{C}}(z,z')$.
Here, $\Delta^F_\sigma(z,z'; z_p) = J^{*}G^F_{\bar{\sigma}}(z,z'; z_p)J^*$ is the hybridization function, which follows from the lattice self-consistency condition for the infinitely coordinated Bethe lattice ($\bar{\sigma}$ denotes the spin opposite to $\sigma$, i.e., $\bar{\sigma} =\ \downarrow$ if $\sigma =\ \uparrow$).
Note that we use the symmetry between the A and B sublattices, i.e., $G^F_{B\sigma}(z,z'; z_p) = G^F_{A\bar{\sigma}}(z,z'; z_p)$~\cite{tsuji2013b}.
Owing to this symmetry, it suffices to solve the effective impurity model only for the A sublattice. In the following, we focus on the A sublattice and suppress the sublattice index.
$\Sigma^{F,{\rm int}}_{\sigma}$ is the interaction self-energy of the impurity model and
$\delta\Sigma^{\rm ext}_\sigma$ is the self-energy due to the external probe field.

We approximate the impurity self-energy with second-order perturbation theory, retaining only the Hartree-Fock and second-order Born contributions~\cite{picano2021},
i.e., $\Sigma^{F,{\rm int}}_{\sigma} = \Sigma^{F,{\rm HF}}_\sigma + \Sigma^{F,{\rm 2B}}_\sigma$.
The self-energies are written as 
\begin{align}
  \Sigma^{F,{\rm HF}}_\sigma(z,z' ; z_p) &= U(z)\left(n^F_{\bar{\sigma}}(z; z_p) - \frac{1}{2}\right) \delta_{\mathcal{C}}(z,z'), \\
  \Sigma^{F,{\rm 2B}}_\sigma(z,z' ; z_p) &= U(z)G^F_{\bar{\sigma}}(z,z'; z_p)G^F_{\bar{\sigma}}(z',z; z_p)G^F_{\sigma}(z,z'; z_p)U(z'). 
\end{align}
Here, $n^F_{\sigma}(z; z_p)\equiv -i G^F_{\sigma}(z,z+0_{\mathcal{C}}; z_p) $ is the occupation of spin $\sigma$
($z+0_{\mathcal{C}}$ means a value infinitesimally larger than $z$ on the contour $\mathcal{C}$).
The self-energy due to the external probe field is
\begin{align}
  \delta\Sigma^{\rm ext}_\sigma(z,z'; z_p) = \frac{\delta h (-1)^{\sigma}}{2} \delta_{\mathcal{C}}(z,z_p) \delta_{\mathcal{C}}(z,z'),
\end{align}
where $(-1)^{\sigma} \equiv -1$ if $\sigma = \uparrow$ and $(-1)^{\sigma} \equiv 1$ if $\sigma = \downarrow$.

In this work, we calculate the spin-spin correlation function as the response of the staggered magnetization $\hat{B}=\hat{m}$ when a staggered magnetic field is applied.
The external field couples to $\hat{A} = -\sqrt{N}\hat{S}^z_{\pi}$ and the spin-spin correlation function is defined as $\chi(z, z_p) \equiv -\delta \langle \hat{m}(z)\rangle_{z_p}/\delta h$.
The quantity $\delta \langle \hat{m}(z)\rangle_{z_p}/\delta h$ can be directly calculated from $G^{\prime}_{\sigma} \equiv \delta G_\sigma / \delta h$ with $\delta G_\sigma = G^F_{\sigma} - G_{\sigma}$, 
so we focus on the calculation of $G^{\prime}_{\sigma}$ in the following.

Unlike in the general formulation of the main text, in DMFT the hybridization function $\Delta_\sigma$ is itself a functional of the local Green's function through the lattice self-consistency condition, so its variation also contributes to the response.
Equation~\eqref{eq:perturbative-GCH} thus applies to the effective impurity model with the replacement $\Sigma^{\rm int}_{\sigma} \to \Delta_{\sigma} + \Sigma^{\rm int}_{\sigma}$, whose kernel contains the additional contribution $\delta \Delta_{\sigma}(z_1,z_2)/\delta G_{\bar{\sigma}}(z_3,z_4)$.
Using
$\Delta^{\prime}_{\sigma} \equiv \delta \Delta_\sigma / \delta h$ with $\delta \Delta_\sigma = \Delta^F_{\sigma} - \Delta_{\sigma}$
and $\Sigma^{\prime\rm int}_{\sigma} \equiv \delta \Sigma^{\rm int}_\sigma / \delta h$ with $\delta \Sigma^{\rm int}_\sigma = \Sigma^{F,{\rm int}}_{\sigma} - \Sigma^{\rm int}_{\sigma}$,
the small changes in the hybridization function and self-energy are written as
\begin{align}
  \Delta^{\prime}_{\sigma}(z_1,z_2; z_p) &= J^{*} G^{\prime}_{\bar{\sigma}}(z_1,z_2; z_p) J^*, \label{eq:sm-hyb-prime}\\
  \Sigma^{\prime,{\rm HF}}_{\sigma}(z_1,z_2; z_p)&= \left\{-iU(z_1)\delta_{\mathcal{C}}(z_1,z_2)\right\}G^{\prime}_{\bar{\sigma}}(z_1,z_1+0_\mathcal{C}; z_p), \label{eq:sm-sigma-hf-prime}\\
  \Sigma^{\prime,{\rm 2B}}_{\sigma}(z_1,z_2; z_p)&= U(z_1)G_{\bar{\sigma}}(z_1,z_2)G_{\bar{\sigma}}(z_2,z_1)U(z_2)G^{\prime}_{\sigma}(z_1,z_2; z_p) \nonumber \\
  &~+ U(z_1)G_\sigma(z_1,z_2)G_{\bar{\sigma}}(z_2,z_1)U(z_2)G^{\prime}_{\bar{\sigma}}(z_1,z_2; z_p) \nonumber \\
  &~+ U(z_1)G_\sigma(z_1,z_2)G_{\bar{\sigma}}(z_1,z_2)U(z_2)G^{\prime}_{\bar{\sigma}}(z_2,z_1; z_p). \label{eq:sm-sigma-2b-prime}
\end{align}
In our calculation, we do not use the kernel $K$ explicitly.
Nevertheless, it can be written as
\begin{align}
  K_{\sigma\sigma}(z_1,z_2; z_3, z_4)
  \equiv \frac{\delta\Sigma^{\rm int}_\sigma(z_1,z_2)}{\delta G_\sigma(z_3,z_4)}
  &=U(z_1)\delta_{\mathcal{C}}(z_1,z_3)G_{\bar{\sigma}}(z_1,z_2)G_{\bar{\sigma}}(z_2,z_1)U(z_2)\delta_{\mathcal{C}}(z_2,z_4), \\
  K_{\sigma\bar{\sigma}}(z_1,z_2; z_3, z_4)
  \equiv \frac{\delta\Sigma^{\rm int}_\sigma(z_1,z_2)}{\delta G_{\bar{\sigma}}(z_3,z_4)}
  &=-iU(z_1)\delta_{\mathcal{C}}(z_1,z_2)\delta_{\mathcal{C}}(z_1,z_3)\delta_{\mathcal{C}}(z_3,z_4) \nonumber \\
  &~+U(z_1)\delta_{\mathcal{C}}(z_1,z_3)G_\sigma(z_1,z_2)G_{\bar{\sigma}}(z_4,z_3)U(z_2)\delta_{\mathcal{C}}(z_2,z_4) \nonumber \\
  &~+U(z_1)\delta_{\mathcal{C}}(z_1,z_4)G_\sigma(z_1,z_2)G_{\bar{\sigma}}(z_4,z_3)U(z_2)\delta_{\mathcal{C}}(z_2,z_3). \label{eq:sm-2b-kernel}
\end{align}
Owing to the contour delta functions, the remaining coefficient in each term is a two-time function, so this kernel could be stored in $\mathcal{O}(N_t^2)$ memory and applied by element-wise products.
This is the sparse structure noted in the End Matter.
It is specific to such low-order approximations and is absent for self-energies involving internal time integrations, such as the $GW$ approximation, where $K$ becomes a dense four-time function.

The linear integral equation for $G^{\prime}_{\sigma}$ is written as
\begin{align}
    [\mathcal{A} \star G^{\prime}_{\sigma}](z,z'; z_p) = \mathcal{B}_{\sigma} (z,z'; z_p).
\end{align}
The left-hand side is written as
\begin{align}
    [\mathcal{A} \star G^{\prime}_{\sigma}](z,z'; z_p) &= G^{\prime}_{\sigma}(z,z'; z_p) - \left[G_{\sigma} * \left(\Delta^{\prime}_{\sigma} + \Sigma^{\prime\rm int}_{\sigma} \right) * G_{\sigma}\right](z, z'; z_p).
\end{align}
The right-hand side is written as
\begin{align}
    \mathcal{B}_{\sigma} (z,z'; z_p) = \frac{(-1)^{\sigma}}{2} G_{\sigma}(z, z_p)G_{\sigma}(z_p, z').
\end{align}
Note that the Hamiltonian has particle-hole symmetry, and the relations $G^F_{\sigma}(z\in \mathcal{C}_i, z' \in \mathcal{C}_j; z_p) = -G^F_{\bar{\sigma}}(z' \in \mathcal{C}_j, z \in \mathcal{C}_i; z_p)$ and $\delta G_{\sigma}(z\in \mathcal{C}_i, z' \in \mathcal{C}_j; z_p) = -\delta G_{\bar{\sigma}}(z' \in \mathcal{C}_j, z \in \mathcal{C}_i; z_p)$ hold~\cite{tsuji2013b}.
Therefore, in our numerical calculations we exploit this symmetry and solve the linear integral equation only for $G^{\prime}_{\uparrow}$.

When we neglect vertex corrections, the kernel $K$ vanishes, which corresponds to dropping $\Sigma^{\prime,{\rm HF}}_{\sigma}$ and $\Sigma^{\prime,{\rm 2B}}_{\sigma}$ and retaining only $\Delta^{\prime}_{\sigma}$.
In the analysis of Fig.~\ref{fig:neq-order-fluctuation}(d), we include only the Hartree-Fock term in the kernel $K$, i.e., $\Sigma^{\prime,{\rm HF}}_{\sigma}$.
Note that in this analysis, the one-particle Green's function $G_{\sigma}$ is still evaluated using the self-energy up to second order, retaining the Hartree-Fock and second-order Born contributions.

\section{Estimation of the Higgs-mode frequency}\label{sm:check-higgs-mode}
\begin{figure*}[hbt]
  \includegraphics[width=\textwidth]{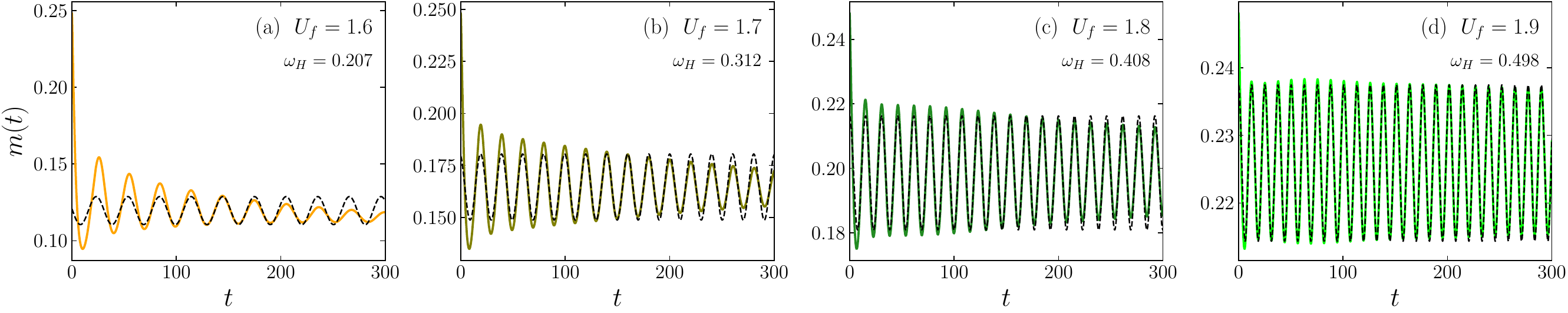}
  \caption{
    Comparison between the order parameter $m(t)$ (solid lines) and trigonometric functions with the estimated Higgs-mode frequency $\omega_H$ (black dashed lines) for (a) $U_f=1.6$, (b) $1.7$, (c) $1.8$, and (d) $1.9$.
  }
  \label{fig:neq-oscillation-fit}
\end{figure*}

In Fig.~\ref{fig:nonequilibrium-critical}(a),
we estimate the Higgs-mode frequency $\omega_H$ from the Fourier transform of the order parameter $m(t)$ multiplied by a Gaussian window function $\propto \exp\left[-(t-t_c)^2/(2w^2)\right]$ centered at $t_c = 150$ with width $w = 30$ for $U_f=1.6, 1.7, 1.8$, and $1.9$.
Figure~\ref{fig:neq-oscillation-fit} compares the order parameter $m(t)$ with trigonometric functions at the frequency $\omega_H$ for $U_f=1.6, 1.7, 1.8$, and $1.9$.
The estimated frequency $\omega_H$ is consistent with the oscillation of the order parameter $m(t)$.
For $U_f = 1.5$, the oscillation of $m(t)$ is strongly damped and only the first few oscillation periods are visible, so we instead estimate the frequency from the positions of the peaks and valleys of $m(t)$ ($\omega_H = 0.102$).

\section*{References}
The references cited in this Supplemental Material are listed in the bibliography of the main text.

\end{document}